\documentclass{article}%
\usepackage{amsfonts}
\usepackage{amsmath}
\usepackage{amssymb}
\usepackage{graphicx}%
\providecommand{\U}[1]{\protect\rule{.1in}{.1in}}
\begin{document}

\title{On effective mass of a photon in a strong magnetic field }
\author{V.M. Katkov\\Budker Institute of Nuclear Physics,\\Novosibirsk, 630090, Russia\\e-mail: katkov@inp.nsk.su}
\maketitle

\begin{abstract}
For the magnetic field in order of the Schwinger critical value or much larger
it, the effective mass of a real photon with a preset polarization is
investigated in the energy region including two lower creation thresholds of
electron and positron on Landau levels. In the high-energy range, when the
number of thresholds is large, the quasiclassical approach is used.

\end{abstract}

\textbf{1}. In 1971, Adler \cite{[1]} had calculated the photon polarization
operator in a constant and homogenous magnetic field using the proper-time
technique developed by Schwinger \cite{[2]}. The polarization operator on mass
shell ( $k^{2}=0,$the metric $ab=$ $a^{0}b^{0}-\mathbf{ab}$ is used ) in a
strong magnetic field $H\gtrsim H_{0}=m^{2}/e=4,4\cdot10^{13}$ $%
%TCIMACRO{\unit{G}}%
%BeginExpansion
\operatorname{G}%
%EndExpansion
$ (the system of units $\hbar=c=1$ is used ) was investigated well enough in
the energy region lower the pair creation threshold (see, for example, the
papers \cite{[3]}, \cite{[4]} and the bibliography cited there). Here we
consider the polarization operator for energies less than the third creation
threshold of electron and positron on Landau levels. We investigate also the
effective mass in the region of large thresold number using the quasiclassical
approach.The general case of an arbitrary value of the photon energy and
magnetic field strength, we shall consider in another work.

Our analysis is based on the general expression for the contribution of spinor
particles to the polarization operator obtained in a diagonal form in
\cite{[5]} (see Eqs. (3.19), (3.33)). For the case of pure magnetic field we
have in a covariant form the following expression%

\begin{align}
\Pi^{\mu\nu}  &  =-\sum_{i=2,3}\kappa_{i}\beta_{i}^{\mu}\beta_{i}^{\nu
},\ \ \ \beta_{i}\beta_{j}=-\ \delta_{ij},\ \ \ \beta_{i}k=0;\label{1}\\
\beta_{2}^{\mu}  &  =(F^{\ast}k)^{\mu}/\sqrt{-(F^{\ast}k)^{2}},\ \ \ \beta
_{3}^{\mu}=(Fk)^{\mu}/\sqrt{-(F^{\ast}k)^{2}},\ \nonumber\\
\ \mathrm{Tr}FF^{\ast}  &  =0,\ \ \mathrm{Tr}F^{2}=F^{\mu\nu}F_{\mu\nu
}=2(H^{2}-E^{2})\equiv2f>0, \label{01}%
\end{align}
where $F^{\mu\nu}-$ the electromagnetic field tensor , $F^{\ast\mu\nu}-$ dual
tensor, $k^{\mu}$ $-$ the photon momentum, $(Fk)^{\mu}=F^{\mu\nu}k_{\nu},$%

\begin{equation}
\kappa_{i}=\frac{\alpha}{\pi}m^{2}r%
%TCIMACRO{\tint \limits_{-1}^{1}}%
%BeginExpansion
{\textstyle\int\limits_{-1}^{1}}
%EndExpansion
dv%
%TCIMACRO{\tint \limits_{0}^{\infty-\mathrm{i0}}}%
%BeginExpansion
{\textstyle\int\limits_{0}^{\infty-\mathrm{i0}}}
%EndExpansion
f_{i}(v,x)\exp[\mathrm{i}\psi(v,x)]dx. \label{2}%
\end{equation}
Here%

\begin{align}
f_{2}(v,x)  &  =2\frac{\cos(vx)-\cos x}{\sin^{3}x}-\frac{\cos(vx)}{\sin
x}+v\frac{\cos x\sin(vx)}{\sin^{2}x},\nonumber\\
f_{3}(v,x)  &  =\frac{\cos(vx)}{\sin x}-v\frac{\cos x\sin(vx)}{\sin^{2}%
x}-(1-v^{2})\cot x,\nonumber\\
\psi(v,x)  &  =\frac{1}{\mu}\left\{  2r\frac{\cos x-\cos(vx)}{\sin
x}+[r(1-v^{2})-1]x\right\}  ;\label{3}\\
r  &  =-(F^{\ast}k)^{2}/4m^{2}f,\ \ \ \mu^{2}=f/H_{0}^{2}. \label{03}%
\end{align}
The real part of $\kappa_{i}$ determines the refractive index $n_{i}$ of the
photon with polarization $e_{i}=\beta_{i}$:%

\begin{equation}
\ \ \ \ n_{i}=1-\frac{\mathrm{\operatorname{Re}}\kappa_{i}}{2\omega^{2}}.
\label{4}%
\end{equation}
At $r>1$, the proper value of polarization operator $\kappa_{i}$ includes the
imaginary part which determines the probability per unit length of pair
production by photon with the polarization $\beta_{i}$:%

\begin{equation}
W_{i}=-\frac{1}{\omega}\mathrm{\operatorname{Im}}\kappa_{i} \label{5}%
\end{equation}
For $r<1$, the integration counter over $x$ in Eq. (\ref{2}) can be turn to
the lower semiaxis $(x\rightarrow-\mathrm{i}x),$ then the value $\kappa_{i}$
becomes real in an explicit form.

\textbf{2. }At $r<1,$ the expression for $\kappa_{i}$ takes the following form:%

\begin{equation}
\kappa_{i}=\alpha m^{2}\frac{r}{\pi}%
%TCIMACRO{\tint \limits_{-1}^{1}}%
%BeginExpansion
{\textstyle\int\limits_{-1}^{1}}
%EndExpansion
dv%
%TCIMACRO{\tint \limits_{0}^{\infty}}%
%BeginExpansion
{\textstyle\int\limits_{0}^{\infty}}
%EndExpansion
F_{i}(v,x)\exp[-\chi(v,x)]dx,\ \ \label{141}%
\end{equation}
Here%

\begin{equation}
F_{2}(v,x)=\frac{1}{\sinh x}\left(  2\frac{\cosh x-\cosh(vx)}{\sinh^{2}%
x}-\cosh(vx)+v\sinh(vx)\coth x\right)  , \label{142}%
\end{equation}

\begin{equation}
F_{3}(v,x)=\frac{\cosh(vx)}{\sinh x}-v\frac{\cosh x\sinh(vx)}{\sinh^{2}%
x}-(1-v^{2})\coth x; \label{143}%
\end{equation}

\begin{equation}
\chi(v,x)=\frac{1}{\mu}\left[  2r\frac{\cosh x-\cosh(vx)}{\sinh x}%
+(rv^{2}-r+1)x\right]  . \label{144}%
\end{equation}
For the energy sufficiently close to the thresold ($(1-r)/\mu\ll1$), we add to
the integrand for $\kappa_{3}$ in (\ref{141}) and take off the function%

\begin{equation}
(1-v^{2})\exp[-\chi_{00}(v,x)],\ \ \chi_{00}(v,x)=\frac{1}{\mu}\left[
2r+(rv^{2}-r+1)x\right]  . \label{145}%
\end{equation}
Integrating over $x$ the deducted part of the integrand, we have%

\begin{equation}
\kappa_{3}^{00}=-\alpha m^{2}\frac{\mu}{\pi}\exp\left(  -\frac{2r}{\mu
}\right)
%TCIMACRO{\tint \limits_{-1}^{1}}%
%BeginExpansion
{\textstyle\int\limits_{-1}^{1}}
%EndExpansion
dv\frac{r(1-v^{2})}{rv^{2}-r+1}. \label{245}%
\end{equation}
After integration over $v,$ we recover $\kappa_{3}$ in the following
well-behaved form:%

\begin{align}
\kappa_{3}  &  =\kappa_{3}^{1}+\kappa_{3}^{00},\ \ \kappa_{3}^{1}=\alpha
m^{2}\frac{r}{\pi}\nonumber\\
&  \times%
%TCIMACRO{\tint \limits_{-1}^{1}}%
%BeginExpansion
{\textstyle\int\limits_{-1}^{1}}
%EndExpansion
dv%
%TCIMACRO{\tint \limits_{0}^{\infty}}%
%BeginExpansion
{\textstyle\int\limits_{0}^{\infty}}
%EndExpansion
\left\{  F_{3}(v,x)\exp[-\chi(v,x)]+(1-v^{2})\exp[-\chi_{00}(v,x)]\right\}
dx,\label{146}\\
\kappa_{3}^{00}  &  =\alpha m^{2}\frac{\mu}{\pi}\exp\left(  -\frac{2r}{\mu
}\right)  \left[  2+B(r)\right]  ;\label{147}\\
\ B(r)  &  =\frac{2}{\sqrt{r(1-r)}}\arctan\sqrt{\frac{1-r}{r}}-\frac{\pi
}{\sqrt{r(1-r)}}\ . \label{148}%
\end{align}
For superstrong fields ($\mu\gg1)$, the value $x\lesssim1$ contributes in the
integral for $\kappa_{2}$ and $\kappa_{3}^{1}$ and the exponential terms in
the integrands can be substitute for unit. As a result, we have for the
leading terms of expansion in series of $\mu$:%

\begin{equation}
\kappa_{2}\simeq-\frac{4r}{3\pi}\alpha m^{2},\ \ \ \ \kappa_{3}\simeq\alpha
m^{2}\frac{\mu}{\pi}(2+B(r)). \label{150}%
\end{equation}
Near the threshold, when $1-r\ll1$ , $B(r)\simeq2-\pi/\sqrt{1-r},$ and we obtain:%

\begin{equation}
\kappa_{2}\simeq-\frac{4}{3\pi}\alpha m^{2},\ \ \ \ \kappa_{3}\simeq-\alpha
m^{2}\mu\frac{1}{\sqrt{1-r}}\left(  1-\frac{4}{\pi}\sqrt{1-r}\right)  .
\label{151}%
\end{equation}
In low-energy range ($r\ll1$), we have:%

\begin{equation}
B(r)\simeq-2-\frac{4}{3}r,\ \ \kappa_{2}\simeq-\frac{4r}{3\pi}\alpha
m^{2},\ \ \kappa_{3}\simeq-\frac{4r\mu}{3\pi}\alpha m^{2}. \label{152}%
\end{equation}
Eq. (\ref{152}) coinsides with Eqs. (2.4), (2.9) in \cite{[3]}.

\textbf{3}. We go on to the next energy region, which upper boundary is higher
the second threshold $r_{10}$ (but not too close to the third threshold
$r_{20}$). On this threshold, one of the particles is created on the first
excited level and another -- in the ground state. In general case%

\begin{equation}
\ \ r_{lk}=(\varepsilon(l)+\varepsilon(k))^{2}/4m^{2},\ \ \ \varepsilon
(l)=\sqrt{m^{2}+2eHl}=m\sqrt{1+2\mu l}.\ \ \ \label{021}%
\end{equation}
For $1<$ $r<r_{10}$, the integration counter over $x$ in Eq. (\ref{2}) can be
turn to the lower imaginary semiaxis, except the integrand term%

\begin{equation}
-(1-v^{2})\cot x\exp[\mathrm{i}\psi(v,x)]. \label{121}%
\end{equation}
Let's add to Eq. (\ref{121}) and take off the function%

\begin{align}
\mathrm{i}(1-v^{2})\exp[\mathrm{i}\psi_{\mathrm{red}}(v,x)],\ \ \psi
_{\mathrm{red}}(v,x)  &  =\frac{1}{\mu}\left\{  2\mathrm{i}r+[r(1-v^{2}%
)-1]x\right\}  .\label{22}\\
&  \ \ \label{022}%
\end{align}
For the sum of the functions, the integration counter over $x$ can be turn to
the lower semiaxis. For the residuary function, the integral over $x$ has the
following form%

\begin{align}%
%TCIMACRO{\tint \limits_{0}^{\infty}}%
%BeginExpansion
{\textstyle\int\limits_{0}^{\infty}}
%EndExpansion
\exp[\mathrm{i}\psi_{\mathrm{red}}(v,x)]dx  &  =\exp\left(  -\frac{2r}{\mu
}\right)  \frac{\mathrm{i}\mu}{r(1-v^{2})-1+\mathrm{i}0}\nonumber\\
&  =\mu\exp\left(  -\frac{2r}{\mu}\right)  \left[  \mathrm{i}\frac
{\text{\textit{ }}\mathcal{P}}{r-1-rv^{2}}+\pi\delta\left(  r-1-rv^{2}\right)
\right]  . \label{23}%
\end{align}
The operator $\mathcal{P}$ means the principal value integral. Carrying out
the integration over $v,$ we have after not complicated calculations%

\begin{align}
-\mathrm{i}r  &
%TCIMACRO{\tint \limits_{-1}^{1}}%
%BeginExpansion
{\textstyle\int\limits_{-1}^{1}}
%EndExpansion
dv(1-v^{2})\left[  \mathrm{i}\frac{\text{\textit{ }}\mathcal{P}}{r-1-rv^{2}%
}+\pi\delta\left(  r-1-rv^{2}\right)  \right]  =2+B(r),\nonumber\\
B(r)  &  =\frac{2}{\sqrt{r(r-1)}}\ln(\sqrt{r}+\sqrt{r-1})-\frac{\mathrm{i}\pi
}{\sqrt{r(r-1)}}. \label{24}%
\end{align}
Finally the expression for $\kappa_{3}$ takes the form of analytical extension
of Eq. (\ref{146}) into the region $r>1$.

\textbf{4}. The integrals for $\kappa_{2}$ and $\kappa_{3}^{1}$ have the root
divergence at $r=r_{10}.$ To bring out these distinctions in an explicit form,
let's consider the main asymptotic terms of corresponding integrand at
$x\rightarrow\infty$:%

\begin{align}
\kappa_{i}^{10}  &  =\alpha m^{2}r\frac{2}{\pi}%
%TCIMACRO{\tint \limits_{-1}^{1}}%
%BeginExpansion
{\textstyle\int\limits_{-1}^{1}}
%EndExpansion
dv%
%TCIMACRO{\tint \limits_{0}^{\infty}}%
%BeginExpansion
{\textstyle\int\limits_{0}^{\infty}}
%EndExpansion
d_{i}(v)\exp[-\chi_{10}(v,x)]dx,\ \nonumber\\
\ \ d_{2}  &  =v-1,\ d_{3}=1-v-\frac{2r}{\mu}(1-v^{2})\label{28}\\
\chi_{10}(v,x)  &  =\chi_{00}(v,x)+1-v=\frac{2r}{\mu}+\frac{1}{\mu}\left[
(1-v)\mu+rv^{2}-r+1\right]  x. \label{29}%
\end{align}
After elementary integration over $x$, one gets%
\begin{equation}
\kappa_{i}^{10}=\alpha m^{2}\mu r\frac{2}{\pi}\exp\left(  -\frac{2r}{\mu
}\right)
%TCIMACRO{\tint \limits_{-1}^{1}}%
%BeginExpansion
{\textstyle\int\limits_{-1}^{1}}
%EndExpansion
dv\frac{d_{i}(v)}{rv^{2}-\mu v-r+1+\mu}. \label{30}%
\end{equation}
Performing integration over $v$, we have:%

\begin{align}
\kappa_{2}^{10}  &  =\alpha m^{2}\mu r\frac{2}{\pi}\exp\left(  -\frac{2r}{\mu
}\right)  \left[  \frac{\mu/2r-1}{\sqrt{h(r)}}A(r)-\frac{1}{2r}\ln
(2\mu+1)\right]  ,\label{31}\\
\kappa_{3}^{10}  &  =\alpha m^{2}\mu r\frac{2}{\pi}\exp\left(  -\frac{2r}{\mu
}\right) \nonumber\\
&  \times\left[  \frac{\mu/2r-1-2/\mu}{\sqrt{h(r)}}A(r)-\frac{1}{2r}\ln
(2\mu+1)+\frac{2}{\mu}\right]  ,\label{32}\\
A(r)  &  =\arctan\frac{r-\mu/2}{\sqrt{h(r)}}+\arctan\frac{r+\mu/2}{\sqrt
{h(r)}}\nonumber\\
&  =\pi-\arctan\frac{\sqrt{h(r)}}{r-\mu/2}-\arctan\frac{\sqrt{h(r)}}{r+\mu
/2},\label{33}\\
h(r)  &  =(1+\mu)r-r^{2}-\mu^{2}/4. \label{034}%
\end{align}
For $r=r_{10}=(1+\mu+\sqrt{1+2\mu})/2,$ $h(r)=0$ and the values $\kappa
_{i}^{10}$ diverge at $r=r_{10}$:%

\begin{equation}
\kappa_{i}^{10}\simeq-4\alpha m^{2}r\exp\left(  -\frac{2r}{\mu}\right)
\frac{\beta_{i}}{\sqrt{h(r)}},\ \ \beta_{2}=\frac{\mu}{2}-\frac{\mu^{2}}%
{4r},\ \ \beta_{3}=1+\frac{\mu}{2}-\frac{\mu^{2}}{4r}. \label{34}%
\end{equation}
For higher photon energies $r>r_{10},$ a new channel of pair creation arises,
and Eq. (\ref{30}) changes over (cf. (\ref{24})):%

\begin{align}
\kappa_{i}^{10}  &  =\alpha m^{2}\mu r\frac{2}{\pi}\exp\left(  -\frac{2r}{\mu
}\right) \nonumber\\
&  \times%
%TCIMACRO{\tint \limits_{-1}^{1}}%
%BeginExpansion
{\textstyle\int\limits_{-1}^{1}}
%EndExpansion
dvd_{i}(v)\left[  \frac{\mathcal{P}}{rv^{2}-\mu v-r+1+\mu}-\mathrm{i}\pi
\delta(rv^{2}-\mu v-r+1+\mu)\right]  ; \label{35}%
\end{align}
At $r-r_{10}<<1$%

\begin{equation}
\kappa_{i}^{10}\simeq-4\mathrm{i}\alpha m^{2}r\exp\left(  -\frac{2r}{\mu
}\right)  \frac{\beta_{i}}{\sqrt{-h(r)}}. \label{36}%
\end{equation}
This direct procedure of divergence elimination can be extended further.

\textbf{5}. For strong fields and high energy levels ($\mu\gtrsim1,\ r\gg\mu
$), the main contribution to the integral in Eq. (\ref{2}) is given by small
values of $x\sim(\mu/r)^{1/3}<<1.$ Expanding the entering functions Eq.
(\ref{3}) over $x$, and carrying out the change of variable $x=\mu t,$ we get:%

\begin{align}
\kappa_{i}  &  =\frac{\alpha m^{2}\kappa^{2}}{24\pi}%
%TCIMACRO{\tint \limits_{0}^{1}}%
%BeginExpansion
{\textstyle\int\limits_{0}^{1}}
%EndExpansion
\alpha_{i}(v)(1-v^{2})dv%
%TCIMACRO{\tint \limits_{0}^{\infty}}%
%BeginExpansion
{\textstyle\int\limits_{0}^{\infty}}
%EndExpansion
t\exp[-\mathrm{i}(t+\xi\frac{t^{3}}{3})]dt;\ \sqrt{\xi}=\frac{\kappa(1-v^{2}%
)}{4},\nonumber\\
\alpha_{2}  &  =3+v^{2},\ \ \alpha_{3}=2(3-v^{2}),\ \ \kappa^{2}=4r\mu
^{2}=-\frac{(Fk)^{2}}{m^{2}H_{0}^{2}}. \label{15}%
\end{align}
At $\kappa\gg1$ ($\xi\gg1$) the small $t$ contributes to the integral
(\ref{15}) ( $\xi t^{3}\sim1),$ and in the argument of exponent Eq. (\ref{15})
the linear over $t$ term can be omit. The condition $\kappa>>1$ is identically
valid in this case. Carrying out the change of variable:%

\begin{equation}
\xi t^{3}/3=-\mathrm{i}x,\ \ \ t=\exp\left(  \frac{-\mathrm{i}\pi}{6}\right)
\left(  \frac{3x}{\xi}\right)  ^{1/3}, \label{19}%
\end{equation}
one obtains:%

\begin{equation}
\kappa_{i}=\frac{\alpha m^{2}\kappa^{2}}{24\pi}\exp\left(  \frac
{-\mathrm{i}\pi}{3}\right)  \frac{1}{3}\left(  \frac{48}{\kappa^{2}}\right)
^{2/3}\Gamma\left(  \frac{2}{3}\right)
%TCIMACRO{\tint \limits_{0}^{1}}%
%BeginExpansion
{\textstyle\int\limits_{0}^{1}}
%EndExpansion
dv\alpha_{i}(v)(1-v^{2})^{-1/3}. \label{20}%
\end{equation}
After integration over $v$ we have:%

\begin{align}
\kappa_{i}  &  =\frac{\alpha m^{2}(3\kappa)^{2/3}}{7\pi}\frac{\Gamma
^{3}\left(  \frac{2}{3}\right)  }{\Gamma\left(  \frac{1}{3}\right)
}(1-\mathrm{i}\sqrt{3})\beta_{i}\label{21}\\
&  =(\allowbreak0.175\,-0.304\,\allowbreak\mathrm{i})\beta_{i}\alpha
m^{2}\kappa^{2/3},\ \ \ \beta_{2}=1,\ \ \ \beta_{3}=3/2.\nonumber
\end{align}

\textbf{6}. It follows from Eq. (\ref{21}) that for $\alpha\kappa^{2/3}>1,$
the photon effective mass becomes larger than the mass of created electron and
positron. And so, it seems that Eq. (\ref{21}) is valid at the photon energy
fulfilling the condition $\alpha\kappa^{2/3}\ll1.$ It should be noted that
this expression does not depend on the electron mass. At the same time, the
first order of the radiation correction to the electron mass have a form (
$\chi=\varepsilon H_{\perp}/mH_{0},\ \varepsilon\ $is the electron energy):%

\begin{align}
m_{\mathrm{rad}1}^{2}  &  =2D\ \alpha m^{2}\chi^{2/3}=2D\alpha\ \widetilde
{\chi}^{2/3},\ \ D=\frac{7(3)^{1/6}}{27}\Gamma\left(  \frac{2}{3}\right)
(1-\mathrm{i}\sqrt{3})\ \label{210}\\
&  =(\allowbreak0.422\,-0.730\mathrm{i}\,),\ \ \widetilde{\chi}^{2}%
=e^{2}\mathcal{P}F^{2}\mathcal{P,}%
\end{align}
and does not depend on the mass too. The main term in the second order of the
radiaton correction to the mass has a form%

\begin{align}
m_{\mathrm{rad}2}^{2}  &  =\frac{13\alpha^{2}m^{2}\chi}{36\sqrt{3}}\left[
1-\mathrm{i}\frac{2}{\pi}\left(  \ln\frac{\chi}{2\sqrt{3}}-\mathrm{C}%
-\frac{142}{39}\right)  \right] \label{211}\\
&  =0.2085\alpha^{2}m^{2}\chi\lbrack1-0.637\mathrm{i}(\ln\chi-5.461)],
\end{align}
where $\mathrm{C}$ -- Euler's constant. This correction includes the
additional fuctor $\sim\alpha\kappa^{1/3}$ comparing to Eq.\ (\ref{210}), and
formally can be larger the last. But for this value of parameter $\chi,$one
can not use the perturbation theory. In this case instead of the Dirac
equation, we must use the well-known Schwinger equation%

\begin{equation}
\lbrack\widehat{\mathcal{P}}-m-M(\mathcal{P,}F\mathcal{)]}\psi=0,\ \widehat
{\mathcal{P}}=\gamma^{\mu}\mathcal{P}_{\mu}\equiv\gamma\mathcal{P}%
,\ \ \mathcal{P}_{\mu}=\mathrm{i}\frac{\partial}{\partial x^{\mu}}-eA_{\mu
},\ \ \ \label{212}%
\end{equation}
where $M$ is the mass operator including, generally speaking, the all series
of the perturbation theory. At substitution $m_{\mathrm{rad}1}$in place of $m$
into $\alpha\kappa^{1/3},$ we have a value $\sim\sqrt{\alpha}/5$, not
depending on any parameter. This value is small and therefore, we can expect
that the first order of the perturbation theory contributs mainly into the
mass operator $M.$ In this order, the mass operator have a relatively simple form%

\begin{equation}
M_{1}(\mathcal{P,}F\mathcal{)\simeq}D\alpha\widetilde{\chi}^{-4/3}(e^{2}\gamma
F^{2}\mathcal{P).} \label{213}%
\end{equation}
Multiply Eq. (\ref{212}) by the operator $\widehat{\mathcal{P}}+m-M$ and take
into account that%

\[
\left\{  \widehat{\mathcal{P}},e^{2}\gamma F^{2}\mathcal{P}\right\}
=2e^{2}\mathcal{P}F^{2}\mathcal{P=}2\widetilde{\chi}^{2},
\]
and the term $\propto e^{2}\gamma F^{2}\mathcal{P}$ can be omit (that relative
value $\sim$ $m_{\mathrm{rad}1}^{2}/\varepsilon^{2}).$ As a result we have the
following squared equation%

\begin{equation}
(\widehat{\mathcal{P}}^{2}-m^{2}-2D\alpha\ \widetilde{\chi}^{2/3}%
)\psi=(\widehat{\mathcal{P}}^{2}-m^{2}-m_{\mathrm{rad}1}^{2})\psi=0
\label{214}%
\end{equation}
All stated above is valid under condition $\varepsilon eH_{\perp}\gg
(m^{2}+m_{\mathrm{rad}1}^{2})^{3/2}.$ For \ $m_{\mathrm{rad}1}^{2}>m^{2},$
this condition changes over $1\gg\alpha^{3/2}.$

The work was supported by the Ministry of Education and Science of the Russian Federation.

\end{document}